\documentclass[a4paper, 11pt]{article}

\usepackage{fullpage}

\usepackage{mathrsfs}
\usepackage{amsfonts}
\usepackage{amssymb}
\usepackage{amsmath}
\usepackage{amsthm}
\usepackage{graphicx}
\usepackage[usenames,dvipsnames]{color}
\usepackage[colorlinks=true,citecolor=blue,linkcolor=magenta]{hyperref}
\usepackage{ulem}
\usepackage{lmodern}
\usepackage{microtype}
\usepackage{braket}
\usepackage{pifont}
\usepackage{mathtools}
\usepackage{color}
\usepackage{dsfont}
\usepackage{algpascal}
\usepackage{algorithm} 
\usepackage{algpseudocode}
\usepackage{chngcntr}
\usepackage{complexity}
\usepackage{float} 

\newtheorem{theorem}{Theorem}

\newtheorem{definition}{Definition}
\newtheorem{lemma}{Lemma}

\newcommand{\comment}[1]{} 
\newcommand{\idty}[1]{\mathbb{1}} 
\newcommand{\ovsqrt}[1]{\frac{1}{\sqrt{2}}}
\newcommand{\tr}[1]{\mathrm{Tr}}

\begin{document}

\title{Stabiliser states are efficiently PAC-learnable}

\author{Andrea Rocchetto\thanks{University of Oxford and University College London. Email: \href{andrea.rocchetto@spc.ox.ac.uk}{andrea.rocchetto@spc.ox.ac.uk}}}

\date{}
\maketitle

\begin{abstract}
The exponential scaling of the wave function is a fundamental property of quantum systems with far reaching implications in our ability to process quantum information. A problem where these are particularly relevant is quantum state tomography. State tomography, whose objective is to obtain an approximate description of a quantum system, can be analysed in the framework of computational learning theory. In this model, Aaronson (2007) showed that quantum states are Probably Approximately Correct (PAC)-learnable with sample complexity linear in the number of qubits. However, it is conjectured that in general quantum states require an exponential amount of computation to be learned. Here, using results from the literature on the efficient classical simulation of quantum systems, we show that stabiliser states are efficiently PAC-learnable. Our results solve an open problem formulated by Aaronson (2007) and establish a connection between classical simulation of quantum systems and efficient learnability. 
\end{abstract}

\section{Introduction}

The goal of quantum tomography is to produce a description of an unknown quantum state given the ability to perform measurements on the state. It is well known that in order to obtain a complete description of a general $n$-qubit quantum state it is necessary to perform $\Omega\left(\mathrm{exp}(n)\right)$ measurements \cite{haah2017sample, o2016efficient}.

The problem of quantum state tomography has been analysed in the framework of the \textit{Probably Approximately Correct} (PAC) model by Aaronson~\cite{aaronson2007learnability}. Here, a learner tries to predict the outcome of measurements performed on a quantum state given access to a training set of measurement outcomes. In this model it has been proved, and experimentally demonstrated on a photonic platform~\cite{rocchetto2017experimental}, that in order to learn a quantum state it is sufficient to have only $\mathcal{O}(n)$ copies of the state.  However, the proposed learning procedure involves an optimisation problem that, in general, can only be solved in exponential time in the number of qubits. 

Similar hard problems are common in the classical/quantum PAC-learning literature where only some concept classes, like halfspaces~\cite{klivans2004learning} and parity functions~\cite{fischer1992learning, helmbold1992learning}, are known to be efficiently learnable. Hellerstein and Servedio provide an overview of known efficiently learnable classes in~\cite{hellerstein2007pac} while Arunachalam and de Wolf~\cite{arunachalam2017survey} and Ciliberto et al.~\cite{ciliberto2018quantum} review the results at the intersection of learning theory and quantum computation. An important question left open in~\cite{aaronson2007learnability} is whether the same applies to the quantum case and if it is possible to identify classes of states that can be efficiently learned. 

Stabiliser states are a particular class of quantum states that is known to be efficiently simulatable by a classical computer~\cite{gottesman1996class, gottesman1997stabilizer, gottesman1998heisenberg}. Indeed, by making use of a specific family of gates, i.e. the Clifford group, one can show that the evolution of a stabiliser state can be simulated on a classical computer in polynomial time. Although the type of circuits allowed is not powerful enough for universal quantum computation, stabiliser states present a rich variety of properties and play a central role in the theory of error--correction~\cite{gottesman1997stabilizer}. Because these states are highly symmetrical it is possible to construct a representation that grows linearly with $n$. This property makes stabiliser states ideal candidates for the study of efficient learnability. 

In this paper we show that stabiliser states can be learned efficiently under two-outcome Pauli measurements. This solves an open question formulated by Aaronson~\cite{aaronson2007learnability} and establishes an interesting link between what can be efficiently learned and what can be efficiently computed. The proof is simple and involves two stages. First, we construct a state that meets the information-theoretic requirements of PAC-learnability. Second, we show that we can produce predictions of future measurements on the state efficiently. 

The paper is structured as follows. In Section~\ref{sec:lernqstates} we present a PAC-learning model for quantum states and define a criterion for efficient learnability. The stabiliser formalism is introduced in Section~\ref{sec:stab} together with a lemma that characterises the expected values of measurements on stabilisers. In Section~\ref{sec:lernstab}  we prove that stabiliser states are efficiently learnable. We conclude in Section~\ref{sec:conc} where we outline some directions for future work.

\section{Learning quantum states in the PAC model}
\label{sec:lernqstates}

Let us recall some standard definitions in quantum theory. A generic $n$-qubit state $\rho $ is a trace one, positive semidefinite matrix acting on a Hilbert
space of dimension $2^{n}$. Pure states have $\tr{}(\rho^2) =1 $ and the corresponding density matrices are rank-one projectors. Any observation of
a state can be mathematically described by a \textit{positive--operator--valued--measurement} (POVM), $E=\{E^{(j)}\}$, where each $E^{(j)}$ is a Hermitian
positive semidefinite operator such that $\sum_{j}E^{(j)}= I$. The probability of measurement outcome $j$ is $p(j)=$Tr$(E^{(j)}\rho )$. For our purposes, we refer to a measurement of $\rho $ as a ``\textit{two-outcome}" POVM $%
\{E^{(1)}=E,E^{(2)}=I-E\}$ with eigenvalues in $[0,1]$.  

The goal of quantum state tomography is to provide an approximate description of a quantum state given a number of its copies. By introducing a weaker version of tomography, where the goal is to produce an hypothesis state that is hard to distinguish from the true state only with respect to a given distribution over measurements, it is possible to use the technical machinery developed in learning theory  to analyse the information theoretic and computational requirements of state reconstruction problems. A more rigorous formulation of this setting as a learning problem can be made in the following way. Take a set $T$ composed of $m$ measurements and their respective expected values over $\rho$. We assume that the measurements are distributed according to an unknown probability distribution $\mathcal{D}$ over two outcome measurements. For an integer $k$, let $[k]$ denotes the set $\{1,\dots, k\}$. We define $T=\{ (E^{(1)}_i, \operatorname*{Tr}(E^{(1)}_i \rho) ) \}_{i \in [m]}$ as the \textit{training set}. The goal of the learning problem is to predict the expected value of a new measurement $E'$ drawn from $\mathcal{D}$ based on the information contained in $T$. 

A way to formalise this type of learning framework is the PAC model developed by Valiant~\cite{valiant1984theory}. This model has been originally developed for Boolean functions but it has then been extended to real-valued ones by Barlett and Long~\cite{barlett1994fat}. In Valiant's theory a learner tries to approximate with high probability an unknown function $f: \mathcal{X} \rightarrow \mathcal{Y} $ given access to  a training set of $m$ random labelled examples $\{(x_i,f(x_i))\}_{i\in[m]}$. We assume that such a function, often referred to as \textit{target concept}, is part of a class of functions $C =\{c: \mathcal{X} \rightarrow \mathcal{Y} \} $ defined as \textit{concept class}. After processing the training set the learner outputs a hypothesis $h: \mathcal{X} \rightarrow \mathcal{Y}$ that is a good approximation of $f$ with probability $\epsilon$. The parameter $\epsilon$ is called \textit{accuracy parameter} and determines how far the hypothesis $h$, measured according to $\mathcal{D}$, can be from $f$. Because the training set is sampled from a probability distribution we introduce the \textit{confidence parameter} to model the probability of sampling a training set that is not representative of the underlying distribution $\mathcal{D}$.  

We say that a concept class $C$ is \textit{PAC-learnable} if, for every $\mathcal{D}$, $f$, $\delta$, there exists an algorithm $L$ that, when running on $m \geq m_{\mathcal{C}}$ examples generated by $\mathcal{D}$, returns an hypothesis $h$ such that, with probability at least $1-\delta$, 
$$\Pr_{x\sim \mathcal{D}}[h(x)\neq f(x)]\leq\epsilon.$$
Here by $\sim$ we indicate that $x$ is drawn from~$\mathcal{D}$. PAC theory introduces two parameters to classify the efficiency of a learner. The first one, $m_{C}$, is information-theoretic and denotes the minimum number of examples such that there exists an algorithm that PAC-learns the class $C$ requiring at most $m_C$ examples.

We refer to $m_{C}$ as the \textit{sample complexity} of the concept class $C$. The second parameter, the \textit{time complexity}, is computational and corresponds to the runtime of the best learner for the class $C$. We say that a concept class is \textit{efficiently} PAC-learnable if the running time of $L$ is polynomial in $n$, $1/\epsilon$ and $1/\delta$.

The framework of PAC-theory has been used to model a weaker version of quantum state tomography~\cite{aaronson2007learnability}. In this framework, differently from standard tomography, where the task is to approximate the outcome of any expectation value on the state, the goal is to produce an hypothesis able to approximate in high probability only measurements drawn from an unknown probability distribution. More specifically, the learner tries to approximate the expected value  $\operatorname*{Tr} (E ^{(1)} \rho)$ of  a measurement $E^{(1)}$ drawn from an unknown $\mathcal{D}$ given access to a training set $T$. The training set $T$ is composed of $m$ random examples  $T=\{( E_i ^{(1)},\tr{} (E_i ^{(1)} \rho))  \}_{ i\in[m] }$ where $E_i ^{(1)}$ is also drawn from $\mathcal{D}$. Notice that we always take the first element $E_i ^{(1)} $ of each POVM $E_i$. For this reason, in the following, we take $E_i ^{(1)} = E_i$. Based on information contained in the training set the learner outputs a hypothesis $\sigma$ that is used to approximate $\rho$ in the sense of approximation described in the following theorem (corresponding to Theorem $1.1$ in Ref.~\cite{aaronson2007learnability}). In this model the following result holds:

\vspace*{12pt}
\noindent
\begin{theorem} [Learning Theorem]
\label{qoccam}Let $\rho$\ be an $n$-qubit mixed state, let $\mathcal{D}$\ be a
distribution over two-outcome measurements of $\rho$, and let $T= \{( E_i,\tr{} ( E_i \rho))  \}_{ i\in[m] } $\ be a training
set consisting of $m$ measurements drawn independently from
$\mathcal{D}$. \ Also, fix error parameters $\varepsilon,\eta,\gamma,\delta
>0$\ with\ $\gamma\varepsilon\geq7\eta$. \ Call $T$ a
\textquotedblleft good\textquotedblright\ training set if any hypothesis
$\sigma$\ that satisfies%
\begin{equation}
\label{eq:inequalities}
\left\vert \operatorname*{Tr}\left(  E_{i}\sigma\right)  -\operatorname*{Tr}%
\left(  E_{i}\rho\right)  \right\vert \leq\eta
\end{equation}
for all $E_{i}\in T$, also satisfies%
\begin{equation}
\label{eq:islearned}
\Pr_{E\sim \mathcal{D}}\left[  \left\vert \operatorname*{Tr}\left(
E\sigma\right)  -\operatorname*{Tr}\left(  E\rho\right)  \right\vert
>\gamma\right]  \leq\varepsilon.
\end{equation}
Then there exists a constant $K>0$\ such that $T$\ is a good
training set with probability at least $1-\delta$, provided that%
\begin{equation}
\label{eq:samplecomplex}
m\geq\frac{K}{\gamma^{2}\varepsilon^{2}}\left(  \frac{n}{\gamma^{2}%
\varepsilon^{2}}\log^{2}\frac{1}{\gamma\varepsilon}+\log\frac{1}{\delta
}\right)  = m_Q.
\end{equation}
\end{theorem}

\vspace*{12pt}
\noindent
The statement of the theorem can be rephrased in the usual language of the PAC model by introducing the concept class $Q$.

\vspace*{12pt}
\noindent
\begin{definition} \label{def:qconcept}
Let $\mathcal{V}$ be a set of $n$-qubit quantum states and let $\mathcal{M}$ be a set of measurements. Every quantum state $\rho \in \mathcal{V}$ has a corresponding concept $q_{\rho}: \mathcal{M} \rightarrow [0,1]$ where $q_{\rho} (E_i) = \operatorname*{Tr} (E_i  \rho)$ and  $E_i \in \mathcal{M}$. The concept class $Q$ is defined as the set of all concepts $q_{\rho}$ corresponding to quantum states in $\mathcal{V}$:
$$
Q = \{ q_{\rho} \}_{\rho \in \mathcal{V}}.
$$
\end{definition}

\vspace*{12pt}
\noindent
We can state Theorem~\ref{qoccam} in terms of the learnability of the concept class $Q$ in the following way:

\vspace*{12pt}
\noindent
\begin{theorem}
Let $\mathcal{V}$ be a set of $n$-qubit quantum states, let $\mathcal{M}$ be a set of $n$-qubit measurements operators, let $\mathcal{D}$\ be a probability distribution over elements of $\mathcal{M}$, and let $T= \{( E_i,\tr{} ( E_i \rho))  \}_{ i\in[m] } $\ be a training set consisting of $m$ measurements drawn independently from
$\mathcal{D}$. For every quantum state $\rho \in \mathcal{V}$ there exists a concept $q_{\rho} (E_i) = \operatorname*{Tr} (E_i  \rho)$ where  $E_i \in \mathcal{M}$ for all $i$. The concept class $Q = \{ q_{\rho} \}_{\rho \in \mathcal{V}}$ is PAC-learnable. That is, fixed error parameters $\varepsilon,\eta,\gamma,\delta>0$\ with\ $\gamma\varepsilon\geq7\eta$, for every target concept $q_{\rho}$ there exists an algorithm that, with probability $1-\delta$ , when running on  $m\geq m_Q$ examples, returns an hypothesis $q_\sigma$ such that:
$$
\Pr_{E \sim \mathcal{D}}\left[  \left\vert q_\sigma(E) - q_\rho(E) \right\vert
>\gamma\right]  \leq\varepsilon.
$$
\end{theorem}

\vspace*{12pt}
\noindent
Theorem~\ref{qoccam} guarantees that, with an adequate number of examples, any hypothesis that satisfies the optimisation problem in Eq.~\ref{eq:inequalities} will be able to approximately predict a new measurement $E'$ drawn from $\mathcal{D}$ with probability $\epsilon$. This notion can be formalised in the following way:

\vspace*{12pt}
\noindent
\begin{definition} [Condtion for efficient learnability] \label{efflearn}
Let $\mathcal{V}$ be a set of $n$-qubit quantum states, let $\mathcal{M}$ be a set of $n$-qubit measurements operators. The concept class $Q$ is efficiently PAC learnable if, for every target concept $q_\rho : \mathcal{M} \rightarrow [0,1]$ with $\rho\in \mathcal{V}$, fixed an error parameter $\eta>0$, there exists an algorithm $L$ running in $\mathrm{poly}(n,1/\eta)$ that, given a training set $T= \{( E_i,\tr{} ( E_i \rho))  \}_{ i\in[m] } = \{( E_i, q_\rho (E_i))  \}_{ i\in[m] } $ where $m$ respects the condition in Eq.~\ref{eq:samplecomplex}, generates a hypothesis state $\sigma$ that satisfies the following program:
\begin{gather}\label{eq:optcond2}
\forall i \in [m] \quad |\operatorname*{Tr}(E_i  \sigma) - \operatorname*{Tr} (E_i \rho) | \leq \eta,  \\
\sigma \succeq 0,  \quad   \operatorname*{Tr}(\sigma) = 1,   \nonumber  
\end{gather}
where by $\sigma \succeq 0$ we denote the positive semidefiniteness of $\sigma$.
\end{definition}

\vspace*{12pt}
\noindent
In this way the problem of learning quantum states becomes equivalent to solving a semidefinite program. It is known that such problem can be solved efficiently in the dimension of $\sigma$~\cite{hazan2008sparse}. However, the dimension of $\sigma$ scales exponentially with $n$ and thus the optimisation problem is effectively not efficiently computable. Recently, Br\~andao et al. proposed a quantum algorithm that solves this problem efficiently when the measurement matrices have low rank~\cite{brandao2017exponential}. We also note that Theorem~\ref{qoccam} has been recently extended in two directions. First, by Aaronson, to a case where the outcome of every measurement is correctly predicted, with high probability, within a given error, the so called ``shadow tomography''~\cite{aaronson2017shadow}. Second, by Aaronson, Chen, Hazan, and Nayak, to an online and regret-minimisation setting~\cite{aaronson2018online}.

It is important to note that predicting measurement outcomes in the probabilistic setting of PAC theory is not a replacement for standard quantum state tomography. Indeed, because the probability of success in Eq.~\ref{eq:islearned} is measured according to the same $\mathcal{D}$ that provides the examples in the training set, an hypothesis that satisfies the inequalities in Eq.~\ref{eq:inequalities} could be far from the true state in the usual trace distance metric, but hard to distinguish from the true state with respect to the points sampled from $\mathcal{D}$. 

The concept of learning defined in the PAC model is different from the ones that have been adopted in other analyses of the learnability of stabiliser states. For example, in the works of Aaronson and Gottesman~\cite{aaronson2008identifying} and of Montanaro~\cite{montanaro2017learning} the goal of the learner is to identify an unknown stabiliser state using the smallest possible number of its copies. Both approaches can classify an $n$-qubit quantum state using only $O(n)$  of its copies but the algorithm by Montanaro requires measurements to be performed on fewer copies of the state at a time. Similarly, Low focused on determining an unknown element of the Clifford group~\cite{low2009learning} while Zhao, P\'erez--Delgado and Fitzsimons tackled the problem of identifying an unknown graph state~\cite{zhao2016fast}. Note that graph states are a subclass of stabiliser states. A key differences between these approaches and the one we develop here is that in~\cite{aaronson2008identifying,montanaro2017learning,low2009learning,zhao2016fast} the learner can actively choose a set of measurements that maximise the probability of reconstructing the state. In the framework discussed in this paper the goal is not to identify a particular state but to predict the outcome of a measurement randomly sampled from an unknown probability distribution based only on information contained in the training set. 

\section{Stabiliser formalism}

\label{sec:stab}

The Pauli matrices $\{I,X,Y,Z\}$ are a set of Hermitian, idempotent, unitary matrices. Apart from the identity operator the Pauli matrices are traceless. We define the Pauli group $\mathcal{P}_n$ of $n$-qubit as $\mathcal{P}_n = \{\pm 1, \pm i\} \cdot \mathcal \{I, X, Y, Z\}^{\otimes n}$. A general Pauli operator can be written, for example, as $P = X \otimes Z \otimes Z \otimes Y$ but in the following we omit the tensor product signs and write $P = XZZY$. For every $P, Q \in \mathcal{P}_n$ we either have $[P,Q]=0$ or $\{P,Q\}=0$, i.e. either their commutator or their anticommutator is zero.

The Pauli group plays a central role in the theory of stabilisers~\cite{gottesman1996class, gottesman1997stabilizer, gottesman1998heisenberg, garcia2014geometry}.
We say that a vector $\ket{\psi}$ is \textit{stabilised} by $P \in \mathcal{P}_n$ if $P\ket{\psi} = \ket{\psi}$. The vectors stabilised by all the elements of a subgroup $\mathcal{S}$ of $\mathcal{P}_n$ form a subspace $V_\mathcal{S}$. $\mathcal{S}$ is called the \textit{stabiliser} of $V_\mathcal{S}$ whose size is  $|V_\mathcal{S}| = 2^n / |\mathcal{S}|$.  Every vector in $V_\mathcal{S}$ is a \textit{stabiliser state}. When a stabiliser contains $2^n$ elements then $|V_\mathcal{S}| = 1$ and the state stabilised is unique.
 
The only vector stabilised by $-I$ and by two anticommuting operators $P$ or $Q$ is the zero vector (proof: $\ket{\psi} = PQ\ket{\psi}= -QP\ket{\psi} = -\ket{\psi}$).
It is a known fact~\cite{hamermesh2012group} that in order for $\mathcal{S}$ to stabilise a non trivial subspace, then $\mathcal{S}$ must be Abelian and not include $-I$. This implies that $\mathcal{S}$ cannot contain elements with phase $\pm i$ (proof: if $i P \in \mathcal{S}$ then $(i P)^2=-I$).

The \textit{generator} of a group $\mathcal{S}$ is a set of elements $\mathcal{L} = \{ S_1, \dots, S_\ell \} \subseteq \mathcal{S}$  such that every element of $\mathcal{S}$ can be written as a product of (possibly repeated) elements of $\mathcal{L}$. The group generated by the elements of $\mathcal{L}$ is denoted as $\langle \mathcal{L} \rangle = \langle S_1, \dots, S_\ell \rangle = \mathcal{S}$.  For any group, a set of generators is \textit{independent} if removing any generator changes the group generated, $\langle S_1, \dots, S_\ell \rangle \neq \langle S_1, \dots, S_{\ell-1} \rangle$.
It is a known fact from group theory~\cite{hamermesh2012group} that any finite group $\mathcal{G}$ has a generating set of size at most $\mathrm{log}_2 |\mathcal{G}|$. A stabiliser state can be efficiently represented by its generating set.  An important result that makes use of this efficient representation is the Gottesman-Knill theorem~\cite{gottesman1998heisenberg}. The theorem proves that circuits composed by elements of the \textit{normaliser} of the Pauli group, i.e. the Clifford group, can be simulated efficiently on a classical computer. Throughout this paper we consider a circuit acting on a particular class of $n$-qubit quantum states to be classically efficiently simulatable (with respect to a specified class of measurements) when we can compute the probabilities of measurement outcomes by classical circuits to $d$ digits of accuracy in $\mathrm{poly}(n,d)$ time.

The density matrix of every stabiliser state can be expressed in terms of its stabilisers. In order to see that, first note that the operator $(I+S)/2$ when $S$ is a Pauli operator, is a projection onto the $+1$ eigenspace of $S$. Therefore if a stabiliser has generators $S_1, \dots, S_n$ then the density matrix for that state is
\begin{equation}
\label{eq:stabrep}
\rho = \frac{1}{2^n} \prod_{i=1} ^n (I + S_i) = \frac{1}{2^n} \sum_{a_1, \dots, a_n \in \{0,1\} } S_1 ^{a_1} \cdots S_n ^{a_n} = \frac{1}{2^n} \sum_{S \in \mathcal{S}} S.
\end{equation}
When we do not have access to the full generating set but only to a subset $\mathcal{L}$ with dimension $|\mathcal{L}| = \ell < n$ we can still construct the projector to the corresponding subspace as $J = \frac{1}{2^\ell} \prod_{i=1} ^\ell (I + S_i)$. In this case, however, the state is not pure and the density matrix corresponds to the projector up to a normalising constant. We thus get for $\ell<n$:
\begin{equation}
\label{eq:stabrep2}
\rho = \frac{1}{2^n} \prod_{i=1} ^\ell (I + S_i) = \frac{1}{2^n} \sum_{S \in \langle \mathcal{L}\rangle} S.
\end{equation}
We note how this expression is still a valid quantum state because $\tr{}(\rho)=1$ and $\rho \succeq 0$ (proof: $\rho$ is equal to a projector up to a normalising constant).

We now prove an easy but useful lemma. In the following we assume that $\rho$ is a stabiliser state, $P_i$ is a Pauli measurement and $S_i$ a stabiliser of $\rho$. We construct the POVM elements $E_i ^{(1)}$ and $E_i ^{(2)}$ of the observable $P_i$ by noting that $ E_i ^{(1)} + E_i ^{(2)} = I$ and $ E_i ^{(1)} - E_i ^{(2)}= P_i$. The POVM element $E_i ^{(1)}$ can be then written as $  E_i ^{(1)} = (I + P_i)/2$. Because we always take the first element $E_i ^{(1)}$ of each POVM in the following we take $  E_i ^{(1)} = E_i$ and denote $E_i$ as the POVM associated to $P_i$.

\vspace*{12pt}
\noindent
\begin{lemma}
\label{lemma:mesoutcome}
Let $E=(I + P)/2$ be a POVM measurement associated to a Pauli operator $P$ and $\rho$ an $n$-qubit stabiliser state then $\tr{}(E\rho)$ can only take the following values $\{0,1/2,1\}$ and: 
$$
\left\{
        \begin{array}{l}
            \text{if } \operatorname*{Tr} (E\rho) = 1 \text{ then } P \text{ is a stabiliser of } \rho; \\
            \text{if } \operatorname*{Tr} (E\rho) = 1/2 \text{ then neither } P \text{ nor } -P \text{ is a stabiliser of } \rho; \\
            \text{if } \operatorname*{Tr} (E\rho) = 0 \text{ then } -P \text{ is a stabiliser of } \rho.
        \end{array}
\right.
$$
\end{lemma}

\vspace*{12pt}
\noindent
\begin{proof}
By using the representation in Eq.~\ref{eq:stabrep} 
we can write $\operatorname*{Tr}(E\rho) = \frac{1}{2^n}  \operatorname*{Tr}\left(\sum_{i=1} ^{2^n}  E S_i \right)$. Recalling that all Pauli matrices are traceless apart from the identity we obtain:
$$
\operatorname*{Tr}(E\rho) = \frac{1}{2^{n+1}} \left( 2^n + \operatorname*{Tr}\left(\sum_{S_i \in \mathcal{S} \setminus I} P S_i \right)\right).
$$
The lemma follows by noting that $S_i^2=I$ and $\tr{}(S_i)=0$ for every $S_i\neq I$ and by observing that because $S_i \neq S_j$ for every $i \neq j$ we can only have at most one non-zero element in the sum.
\end{proof}

\section{Learning stabiliser states}

\label{sec:lernstab}

Consider the following learning task: let $\rho$ be the $n$-qubit quantum state stabilised by a non-trivial stabiliser subgroup $\mathcal{S}$. If $|\mathcal{S}|=2^n$ then $\mathcal{S}$ defines a pure state. If instead $|\mathcal{S}|<2^n$  then $\rho$ is the mixed state in Eq.~\ref{eq:stabrep2}. Given a training set $T= \{( E_i,\tr{}(E_i \rho))  \}_{ i\in[m] }$ drawn from an unknown probability distribution $\mathcal{D}$ the goal of the learner is to predict the expected value of a new measurement $E'$ also drawn from $\mathcal{D}$. We assume $\mathcal{D}$ to be over the set of POVM measurements corresponding to elements of the Pauli group $\mathcal{P}_n$. If the number of examples $m$ respects the conditions set by Theorem~\ref{qoccam}, we are guaranteed that $\rho$ can be PAC-learned. It remains to be determined whether stabiliser states can also be efficiently PAC learned. In the language of PAC-theory, efficiently learning stabiliser states corresponds to proving that the concept class $Q$ is efficiently PAC-learnable when the set of states $\mathcal{V}$ corresponds to the set of stabiliser states and the set of measurements $\mathcal{M}$ corresponds to measurements in the Pauli group $\mathcal{P}_n$. Recall that the notion of efficiency is related to the time complexity of the learning problem. In the following, we prove that stabiliser states are efficiently PAC-learnable:

\vspace*{12pt}
\noindent
\begin{theorem}[Stabiliser states are efficiently PAC-learnable]
\label{th:main}
Let $\mathcal{V}$ be a set of $n$-qubit quantum states, let $\mathcal{M}$ be a set of measurements and let $\mathcal{D}$ be a probability distribution over elements of $\mathcal{M}$.  The concept class $Q = \{ q_{\rho}: \mathcal{M} \rightarrow [0,1] \}_{\rho \in \mathcal{V}}$ is efficiently PAC-learnable with respect to $\mathcal{D}$ when $\mathcal{V}$ is the set of stabiliser states on $n$-qubits and $\mathcal{M}$ is the set of measurements associated to the Pauli group $\mathcal{P}_n$. Similarly, we say that the stabiliser states are efficiently PAC-learnable with respect to the Pauli group.
\end{theorem}

\vspace*{12pt}
\noindent
Our proof is structured in the following way. We begin by constructing a hypothesis state that minimises the program in Eq~\ref{eq:optcond2}. However, this hypothesis contains exponentially many (in the number of qubits $n$) terms and cannot be constructed efficiently. We solve this problem by showing that we can make predictions on the hypothesis without producing the full state. This strategy exploits the group structure of the stabilisers and can be implemented in two algorithms that allow us to predict the expected value of a new measurement. Algorithm~\ref{algo:learn1} constructs a list $\mathcal{L}$ of the generators contained in the training set. Algorithm~\ref{algo:predictions} predicts the value of $\tr{}(E'\rho)$ by checking whether it can be generated by the known generators.

Because all the information required to determine a stabiliser state is contained in its generators we construct the hypothesis $\sigma$ by identifying the generators contained in the training set $T$. Recall that, for every measurement $E_i$ in the training set such that $\operatorname*{Tr}(E_i \rho) = 1$ there is an associated stabiliser element $P_i = 2E_i - I$. In order to identify the generators we make use of two results. Thanks to Lemma~\ref{lemma:mesoutcome} we can identify which measurements, if any, in $T$ correspond to a stabiliser measurement of the state. After the first stabiliser measurement has been identified, and placed on a list $\mathcal{L}$, the algorithm checks whether any new $E_i$ such that $\tr{}(E_i \rho )=1$ can be generated from $\mathcal{L}$. At the end of the process the learner returns a list of independent generators $\mathcal{L} = \{ S_1, \dots, S_l \}$. Based on this information our knowledge of the state can be summarised in the following state:
\begin{equation}
\label{eq:hypothesis}
\sigma = \frac{1}{2^n} \sum_{S_i \in \langle \mathcal{L} \rangle } S_i.
\end{equation}
By using Lemma~\ref{lemma:mesoutcome} it is easy to see how $\sigma$ respects all the inequalities in Eq.~\ref{eq:optcond2}. Because the state is also a normalised projector we have that $\sigma \succeq 0$. Note that a simple sum of the known stabilisers would have also satisfied the inequalities in Eq.~\ref{eq:inequalities} but, in general, it would not be positive semidefinite. 

It remains to be given an efficient algorithm to determine whether a new example is independent of the list of generators $\mathcal{L}$ collected so far. This is necessary to predict the value of $\operatorname*{Tr}(E_i ' \rho)$. We do that below using a variant of the check matrix method described in~\cite{nielsen2010quantum}. With this technique every element of $P \in \mathcal{P}_n$, where $P = P^1 \otimes \cdots \otimes P^n$, is mapped to a $2n + 1$ dimensional row vector $r_P \in \{0,1\}^{2n+1}$. The vector $r_P$ is defined in the following way:
\begin{gather*}
r_P(1) = \left \{
                \begin{array}{ll}
                  0 \quad \text{if}\quad \mathrm{sgn}(P) = +1 \\
                  1 \quad \text{if}\quad \mathrm{sgn}(P) = -1
                \end{array} \right. \\
\forall i\in \{1,\dots, n\} \quad r_P(i) = \left \{
                \begin{array}{ll}
                  0 \quad \text{if}\quad P^i = Z \\
                  1 \quad \text{if}\quad P^i \in \{X,Y\}
                \end{array} \right. \\
\forall i\in \{n+1,\dots, 2n\} \quad r_P(i) = \left \{
                \begin{array}{ll}
                  0 \quad \text{if}\quad P^{i-n} = X \\
                  1 \quad \text{if}\quad P^{i-n} \in \{Y,Z\},
                \end{array}
              \right.                
\end{gather*}
where $\mathrm{sgn}(P')=+1$ if the overall sign of $P^{1} \cdots P ^{n}$ is positive and $\mathrm{sgn}(P')=-1$ otherwise. As an example,
$$
-XYZY \rightarrow r(-XYZY) = \left[1 \: | \: 1\: 1 \: 0 \: 1 \: | \: 0 \: 1\: 1 \: 1 \: \right].
$$
By checking whether the set of unsigned binary vectors $\{r_{S_1},\dots, r_{S_l}\}$ is linearly independent we can determine if the corresponding Pauli operators are also independent. We can use Gaussian elimination to perform this operation at a cost of $\mathcal{O}(n^3)$. Algorithm~\ref{algo:learn1} can be used to produce $\mathcal{L}$.

\begin{algorithm}[H]
\caption{Learning}
\label{algo:learn1}
\textbf{Input:} training set $T=\{(E_i,\operatorname*{Tr}(E_i\rho))\}_{i\in [m]}$ where $E_i = (P_i + I ) /2$\\
\textbf{Output:} list of generators $\mathcal{L}$ contained in $T$ \\
\begin{algorithmic}[1]
\For{$k=1$ to $m$}
\If{$\tr{}(E_k \rho) = 1$ or $\tr{}(E_k \rho) = 0$ and $E_k$ is not generated by $\mathcal{L}$} 
\State add $\tr{}(P_k \rho) P_k $ to $\mathcal{L}$
\EndIf
\EndFor
\end{algorithmic}
\end{algorithm}

Because from the generating set $\mathcal{L}$ we can construct up to $2^l$ elements we cannot write down the full hypothesis state $\sigma$ efficiently. But there is no need to construct this state explicitly. By using a technique developed by Aaronson and Gottesman to keep track the evolution of a row vector~\cite{gottesman1996class, aaronson2004improved} we can make use of the information contained in $\sigma$ using only the generators.

For every new measurement $E'$ we want to determine whether $E'$ commutes with the elements of $\mathcal{L}$ and whether it can be generated by $\mathcal{L}$. Both tasks can be accomplished efficiently using the check-matrix representation~\cite{nielsen2010quantum}. However, because the check matrix representation does not allow us to predict the sign, we are left with determining whether it is the operator $P'$ or $-P'$ that can be generated with the elements of $\mathcal{L}$ (recall that $E' = (P' + I)/2$). This can be accomplished in the following way. Because in the check vector representation matrix multiplication between operators corresponds to addition modulo $2$ and we know that $P'$ is generated by $\mathcal{L}$ we can write:
$$
\sum_{i=1} ^l c_i r_{S_i} = r_{P'}
$$
where $c_i \in \{0, 1\}$ and the addition is done modulo $2$. This corresponds to a system of linear equations that can be solved efficiently. Once we have found the right vector $c$ we can multiply the relevant operators (an efficient algorithm is described in~\cite{aaronson2004improved}) to determine the sign:
\begin{equation}
\label{eq:sign}
\mathrm{sgn}(P') = \mathrm{sgn}(S_1 ^{c_1} \cdots S_n ^{c_n}).
\end{equation}
Algorithm~\ref{algo:predictions} describes how to perform the prediction of the expected value of a new measurement $E'$.

The computational cost of Algorithm~\ref{algo:learn1} and \ref{algo:predictions} is dominated by the cost of determining whether the stabiliser measurements in the training set are linearly independent. In the worst case scenario of a training set composed by $m$ stabiliser measurements, the linear independence must be checked $m$ times at a cost of $n^3$ per operation. Therefore, the overall time complexity of learning stabiliser states is $O(mn^3)$.

Finally, we note that Algorithm~\ref{algo:learn1} and \ref{algo:predictions} are exact in the sense that the difference between the true and predicted value of the expected measurement outcomes is $0$. For this reason there is no $\eta$ dependency in the running time of the learning algorithm. 
\begin{algorithm}[H]
\caption{Predictions}
\label{algo:predictions}
\textbf{Input:} set of known stabiliser generators $\mathcal{L} = \{S_i\}$, new measurement $E'=(I+P')/2$ \\
\textbf{Output:} prediction $\tr{}(E'\rho)$
\begin{algorithmic}[1]
\If{$[P',S_i]=0$ $\forall i $ and $P'$ is generated by $\mathcal{L}$}
	\State solve for $c_i$ equation $\sum_{i=1} ^l c_i r_{S_i} = r_{P'}$ and determine $\mathrm{sgn}(P')$ with Eq.~\ref{eq:sign}
    \If{$\mathrm{sgn}(P')=1$} 
    	\State $\tr{}(E'\rho)=1$ 
    \ElsIf{$\mathrm{sgn}(P')=-1$}
    		\State $\tr{}(E'\rho) = 0$
    \EndIf
\Else
	\State  $\tr{}(E'\rho) = 1/2$
\EndIf
\end{algorithmic}
\end{algorithm}

\section{Conclusion}
\label{sec:conc}

Building on results from the literature on the efficient classical simulation of stabiliser circuits we proved that stabiliser states can be efficiently PAC-learned. Although previous works~\cite{aaronson2008identifying, montanaro2017learning,low2009learning,zhao2016fast} showed that stabilisers can be learned with access to only $O(n)$ copies of the state and polynomial amount of classical computation our results do not require specific measurements to be made on the state and allow us to PAC learn the state under any probability distribution over the measurements.

Because the evolution of stabiliser states can be simulated efficiently on classical devices this work opens new directions in the study of the power of quantum systems: is it possible to establish a connection between what can be efficiently learned and what can be efficiently classically simulated? As previously suggested by Aaronson~\cite{aaronson2007learnability}, it would be interesting to investigate whether match gates~\cite{valiant2001quantum, theral2002classical, jozsa2008matchgates}, a particular class of quantum states that can be efficiently simulated, can also be efficiently learned. More generally, what can be said about the PAC-learnability of computationally tractable states (\textit{i.e.} states whose quantum evolution can be simulated efficiently with classical methods; for a rigorous definition see the work of Van den Nest~\cite{nest2011simulating} and Schwarz and Van den Nest~\cite{schwarz2013simulating})? 

Finally, a recent work by Brand\~ao and Svore~\cite{brandao2016quantum}, whose upper and lower bounds have been subsequently improved by van Apeldoorn et al.~\cite{apeldoorn2017quantum} and Brand\~ao et al.~\cite{brandao2017exponential}, showed that a quantum computer can solve, under certain assumptions, exponentially faster the semidefinite program in Eq.~\ref{eq:optcond2} when the operators $E_i$ are low rank. It would be interesting to investigate whether this result can help to establish a link between states that can be reconstructed efficiently using low rank POVMs and efficient PAC-learnability. 

\section*{Acknowledgements}

I would like to thank Ronald de Wolf for helpful comments and careful reads of the manuscript and Scott Aaronson, Simon Benjamin, Fernando Br\~andao, Toby Cubitt, Carlos Gonz\'alez Guill\'en, Varun Kanade, Ying Li, Simon Perdrix, Fabio Sciarrino, Simone Severini and two anonymous reviewers for helpful comments and suggestions. The author is supported by an EPSRC DTP Scholarship and by QinetiQ Ltd.

\end{document}